\documentclass{amsart}
\usepackage{amssymb}
\usepackage{amsfonts}
\usepackage{amsmath}
\usepackage{lastpage}
\usepackage[utf8]{inputenc}
\usepackage[legalpaper,bookmarks=true,colorlinks=true,linkcolor=blue,citecolor=blue]{hyperref}
\usepackage{graphicx}%
\usepackage{fancyhdr}
\usepackage{color}
\usepackage[mathlines]{lineno}
\usepackage{lscape}
\usepackage{epsfig}
\usepackage{natbib}
\usepackage[]{listings}
\usepackage{geometry}
\usepackage{tgbonum}
\fontfamily{qcr}\selectfont

\newtheorem{theorem}{Theorem}
\theoremstyle{plain}

\newtheorem{definition}{Definition}

\newtheorem{lemma}{Lemma}

\numberwithin{equation}{section}

\begin{document}
\Large

\title[A review on Association concept]{A Review on asymptotic normality of
sums of associated random variables}
\author{$^{(1,2)}$ Gane Samb LO}
\author{$^{(2,3)}$ Harouna SANGARE}
\author{$^{(2)}$ Cheikhna Hamallah Ndiaye}

\begin{abstract}
In this document, we make a round up of the theory of asymptotic normality of sums of associated random variables, in a coherent approach in view of
further contributions for new researchers in the field.\\

\noindent \textbf{Résumé} : Le concept de variables aléatoires associées associées est une généralisation de l'indépendance entre variables
aléatoires. Il devient de plus en plus important dans les probabilités et les applications statistiques de tous les jours.
Dans ce papier, nous introduisons à ce concept et présentons le théorème fondamental de la normalité asymptotique de sommes partielles d'une
suite stationnaire de variables aléatoires associées. Nous utilisons un cadre moderne et cohérent qui sera profitables aux chercheurs débutant dans ce domaine.\\

\noindent $^{(2)}$ H. Sangaré, Cheikhna Hamallah Ndiaye, Gane Samb Lo.\\
LERSTAD, Université Gaston Berger de Saint-Louis, Sénégal\\
$^{(3)}$ Harouna Sangaré : DER MI, Université des Sciences, des Techniques et des Technologies de Bamako, Mali\\
$^{(1)}$ Gane Samb Lo.\\
LERSTAD, Gaston Berger University, Saint-Louis, S\'en\'egal (main affiliation).\newline
LSTA, Pierre and Marie Curie University, Paris VI, France.\newline
AUST - African University of Sciences and Technology, Abuja, Nigeria\\
gane-samb.lo@edu.ugb.sn, gslo@aust.edu.ng, ganesamblo@ganesamblo.net\\
Permanent address : 1178 Evanston Dr NW T3P 0J9,Calgary, Alberta, Canada.\\

\end{abstract}
\maketitle

\section{A brief Reminder on Association}

We then may begin to introduce to the associated random variables concept
which goes back to \cite{lehmann} in the bivariate case.
Notice that we will lessen the notation by putting $k(n)=k$ in the sequel.%
\newline

\noindent The concept of association for random variables generalizes that
of independence and seems to model a great variety of stochastic models.%
\newline

\noindent This property also arises in Physics, and is quoted under the name
of FKG property (\cite{fortuin}), in percolation theory and even in Finance (see\cite{jiazhu}).%
\newline

\noindent The definite definition is given by \cite{esary} as follows.

\begin{definition}
A finite sequence of rv's $(X_{1},...,X_{n})$ are associated when for any
couple of real and coordinate-wise non-decreasing functions $h$ and $g$
defined on $\mathbb{R}^{n}$, we have 
\begin{equation}
Cov(h(X_{1},...,X_{n}),\ \ g(X_{1},...,X_{n}))\geq 0
\end{equation}%
An infinite sequence of rv's are associated whenever all its finite
subsequences are associated.
\end{definition}

We have a few number of interesting properties to be found in (\cite{rao}%
) :

\bigskip

\noindent \textbf{(P1)} A sequence of independent rv's is associated.\newline

\noindent \textbf{(P2)} Partial sums of associated rv's are associated.%
\newline

\noindent \textbf{(P3)} Order statistics of independent rv's are associated.%
\newline

\noindent \textbf{(P4)} Non-decreasing functions and non-increasing
functions of associated variables are associated.\newline

\noindent \textbf{(P5)} Let the sequence $Z_{1},Z_{2},...,Z_{n}$ be
associated and let $(a_{i})_{1\leq i\leq n}$ be positive numbers and $%
(b_{i})_{1\leq i\leq n}$ real numbers. Then the \textit{rv}'s $%
a_{i}(Z_{i}-b_{i})$ are associated.\newline

\noindent As immediate other examples of associated sequences, we may cite
Gaussian random vectors with nonnegatively correlated components (see 
\cite{pitt}) and a homogenuous Markov chain is also associated (\cite{daley}).\newline

\noindent Demimartingales are set from associated centered variables exactly
as martingales are derived from partial sums of centered independent random
variables. We have

\begin{definition}
A sequence of rv's $\{S_{n},n\geq 1\}$ \ in $L^{1}(\Omega ,\mathcal{A},%
\mathbb{P})$ \ is a demimartingale when for any $j\geq 1$, for any
coordinatewise nondecreasing function $g$ \ defined on $\mathbb{R}^{j}$, we
have 
\begin{equation}
\mathbb{E}\bigl((S_{j+1}-S_{j})\ g(S_{1},...,S_{j})\bigr)\geq 0,\ \ j\geq 1.
\label{defmarting}
\end{equation}
\end{definition}

\bigskip Two particular cases should be highlighted. First any martingale is
a demimartingale. Secondly, partial sums $S_{0}=0$, $S_{n}=X_{1}+...+X_{n}$, 
$n\geq 1$, of associated and centered random variables $X_{1},X_{2},...$
form a demimartingale for, in this case, (\ref{defmarting}) becomes : 
\begin{equation*}
\mathbb{E}\left\{ (S_{j+1}-S_{j})\ g(S_{1},...,S_{j})\right\} =\mathbb{E}%
\left\{ X_{j+1}\ g(S_{1},...,S_{j})\right\} =Cov\left\{
X_{j+1},g(S_{1},...,S_{j})\right\} ,
\end{equation*}%
since $\mathbb{E}X_{j+1}=0$. Since $(x_{1},...,x_{j+1})\longmapsto x_{j+1}$
et $(x_{1},...,x_{j+1})\longmapsto g(x_{1},...,x_{j})$ are coordinate-wise
nondecreasing functions and since the $X_{1},X_{2},..$ are associated, we get%
\begin{equation*}
\mathbb{E}\left\{ (S_{j+1}-S_{j})\ g(S_{1},...,S_{j})\right\} =Cov\left\{
X_{j+1}\ g(S_{1},...,S_{j})\right\} \geq 0.
\end{equation*}

\bigskip

\bigskip

\section{Key results for associated sequences}

\begin{lemma}
\label{lemg1} Let $(X,Y)$ be a bivariate random vector such that $\mathbb{E}%
(X^{2})<\infty $ and $\mathbb{E}(Y^{2})<\infty .$ If $\left(
X_{1},Y_{1}\right) $ and $\left( X_{2},Y_{2}\right) $ are two independent
copies of $(X,Y),$ then We have 
\begin{equation*}
2Cov(X,Y)=\mathbb{E}(X_{1}-X_{2})(Y_{1}-Y_{2}).
\end{equation*}%
We also have%
\begin{equation*}
Cov(X,Y)=\int_{-\infty }^{+\infty }\int_{-\infty }^{+\infty }H(x,y)dxdy,
\end{equation*}
\end{lemma}

\noindent where,%
\begin{equation*}
H(x,y)=\mathbb{P}(X>x,Y>y)-\mathbb{P}(X>x)\mathbb{P}(Y>y).
\end{equation*}

\bigskip

\noindent Before the proof of the lemma, we observe that : 
\begin{equation}
H(x,y)=\mathbb{P}(X>x,Y>y)-\mathbb{P}(X>x)\mathbb{P}(Y>y)=\mathbb{P}(X\leq
x,Y\leq y)-\mathbb{P}(X\leq x)\mathbb{P}(Y\leq y).  \label{equiv}
\end{equation}
Indeed we have 
\begin{equation*}
\mathbb{P}(X>x,Y>y)-\mathbb{P}(X>x)\mathbb{P}(Y>y)=\mathbb{E}(\mathbb{I}%
_{\left( X>x\right) }\mathbb{I}_{\left( Y>y\right) })-\mathbb{E}(\mathbb{I}%
_{\left( X>x\right) })\mathbb{E}(\mathbb{I}_{\left( Y>y\right) })
\end{equation*}%
\begin{equation*}
=Cov(\mathbb{I}_{\left( X>x\right) },\mathbb{I}_{\left( Y>y\right) })
\end{equation*}%
\begin{equation*}
=Cov(1-\mathbb{I}_{\left( X>x\right) },1-\mathbb{I}_{\left( Y>y\right) })
\end{equation*}%
\begin{equation*}
=Cov(\mathbb{I}_{\left( X\leq x\right) },\mathbb{I}_{\left( Y\leq y\right) })
\end{equation*}%
\begin{equation*}
=\mathbb{P}(X\leq x,Y\leq y)-\mathbb{P}(X\leq x)\mathbb{P}(Y\leq y).
\end{equation*}

\bigskip

\noindent \textbf{Proof. }We have 
\begin{equation*}
\mathbb{E}(X_{1}-X_{2})(Y_{1}-Y_{2})=\mathbb{E}(X_{1}Y_{1})-\mathbb{E}(X_{1})%
\mathbb{E}(Y_{2})-\mathbb{E}(X_{2})\mathbb{E}(Y_{1})+\mathbb{E}(X_{2}Y_{2})
\end{equation*}%
\begin{equation*}
=2\mathbb{E}(X_{1}Y_{1})-2\mathbb{E}(X_{1})\mathbb{E}(Y_{1})
\end{equation*}%
\begin{equation*}
=2Cov(X_{1},Y_{1}).
\end{equation*}

\bigskip \noindent Next, for $a\in \mathbb{R},$by Fubini's Theorem for
nonegative random variables, 
\begin{equation*}
\int_{a}^{\infty }\int_{a}^{\infty }\mathbb{P}(X>x,Y>y)dxdy=\mathbb{E}%
\int_{a}^{\infty }\int_{a}^{\infty }\mathbb{I}_{\left( X>x\right) }\mathbb{I}%
_{\left( Y>y\right) }dxdy
\end{equation*}%
\begin{equation*}
=\mathbb{E}(\int_{a}^{X}dx\int_{a}^{Y}dy)
\end{equation*}%
\begin{equation*}
=\mathbb{E}[(X-a)(Y-a)].
\end{equation*}

\bigskip \noindent \noindent We have 
\begin{equation*}
2Cov(X_{1},Y_{1})=\mathbb{E}(X_{1}-X_{2})(Y_{1}-Y_{2})
\end{equation*}%
\begin{equation*}
=\mathbb{E}(\left\{ (X_{1}-a)-(X_{2}-a)\right\} \left\{
(Y_{1}-a)-(Y_{2}-a)\right\} ))
\end{equation*}%
\begin{equation*}
=\mathbb{E}(X_{1}-a)(Y_{1}-a)-\mathbb{E}(X_{1}-a)(Y_{2}-a)
\end{equation*}%
\begin{equation*}
-\mathbb{E}(X_{2}-a)(Y_{1}-a)+\mathbb{E}(X_{2}-a)(Y_{2}-a)
\end{equation*}%
\begin{equation*}
=\int_{a}^{\infty }\int_{a}^{\infty }\mathbb{P}(X_{1}>x,Y_{1}>y)dxdy-%
\int_{a}^{\infty }\int_{a}^{\infty }\mathbb{P}(X_{1}>x,Y_{2}>y)dxdy
\end{equation*}%
\begin{equation*}
-\int_{a}^{\infty }\int_{a}^{\infty }\mathbb{P}(X_{2}>x,Y_{1}>y)dxdy+%
\int_{a}^{\infty }\int_{a}^{\infty }\mathbb{P}(X_{2}>x,Y_{2}>y)dxdy.
\end{equation*}

\bigskip \noindent By the independence of $\{X_{1},Y_{1}\}$ and $%
\{X_{2},Y_{2}\},$ $\mathbb{P}(X_{1}>x,Y_{2}>y)=\mathbb{P}(X_{1}>x)\times 
\mathbb{P}(Y_{1}>y)$ and $\mathbb{P}(X_{2}>x,Y_{1}>y)=\mathbb{P}%
(X_{1}>x)\times \mathbb{P}(Y_{1}>y),$%
\begin{equation*}
2Cov(X,Y)=2\left( \int_{a}^{\infty }\int_{a}^{\infty }\left\{ \mathbb{P}%
(X_{1}>x,Y_{1}>y)-\mathbb{P}(X_{1}>x)\times \mathbb{P}(Y_{1}>y)\right\}
dxdy\right) .
\end{equation*}

\bigskip

\noindent We get the final result by letting $a\rightarrow -\infty .$

\begin{lemma}
\label{lemg2} Suppose that $X$, $Y$ are two random variables with finite
variance and, $f$ and $g$ are $C^{1}$ complex valued functions on $\mathbb{R}%
^{1}$ with bounded derivatives $f^{\prime }$ and $g^{\prime }.$ Then 
\begin{equation*}
|Cov(f(X),h(Y))|\leq ||f^{\prime }||_{\infty }||g^{\prime }||_{\infty
}Cov(X,Y)
\end{equation*}
\end{lemma}

\bigskip

\noindent \textbf{Proof.} By Lemma \ref{lemg1}, we have 
\begin{equation*}
2Cov(f(X),g(Y))=\mathbb{E}(f(X_{1})-f(X_{2}))(g(Y_{1})-g(Y_{2}))
\end{equation*}%
\begin{equation*}
=\mathbb{E}\left( \int_{X_{1}}^{X_{2}}f^{\prime
}(x)dx\int_{Y_{1}}^{Y_{2}}g^{\prime }(x)dx\right) .
\end{equation*}%
But%
\begin{equation*}
\int_{X_{1}}^{X_{2}}f^{\prime }(x)dx=\int_{X_{1}}^{+\infty }f^{\prime
}(x)dx-\int_{X_{2}}^{+\infty }f^{\prime }(x)dx
\end{equation*}%
\begin{equation*}
=\int_{\mathbb{R}}f^{\prime }(x)\left\{ 1_{(X_{1}\leq x)}-1_{(X_{2}\leq
x)}\right\} dx
\end{equation*}%
Applying this to $\int_{Y_{1}}^{Y_{2}}g^{\prime }(x)dx$ and combining all
that, leads to%
\begin{equation}
2Cov(f(X),g(Y))=\mathbb{E}\int_{\mathbb{R}^{2}}f^{\prime }(x)g^{\prime }(y)\left\{
1_{(X_{1}\leq x)}-1_{(X_{2}\leq x)}\right\} \left\{ 1_{(Y_{1}\leq
y)}-1_{(Y_{2}\leq y)}\right\} dxdy.  \label{int1}
\end{equation}%
It is easy to see that%
\begin{equation*}
\mathbb{E}\left\{ 1_{(X_{1}\leq x)}-1_{(X_{2}\leq x)}\right\} \left\{
1_{(Y_{1}\leq y)}-1_{(Y_{2}\leq y)}\right\}
\end{equation*}%
\begin{equation*}
=2(\mathbb{P}(X\leq x,Y\leq y)-\mathbb{P}(X\leq x)\mathbb{P}(Y\leq y))
\end{equation*}%
and by (\ref{equiv}), this is equal to $2H(x,y).$ By applying Fubini's
theorem in (\ref{int1}), we get%
\begin{equation*}
2Cov(f(X),g(Y))=2\int_{\mathbb{R}^{2}}f^{\prime }(x)g^{\prime }(y)H(x,y)dxdy.
\end{equation*}%
This gives, since $H(x,y)\geq 0$ for associated $rv$'s, 
\begin{equation*}
\left\vert Cov(f(X),g(Y))\right\vert \leq ||f^{\prime }||_{\infty
}||g^{\prime }||_{\infty }\int_{\mathbb{R}^{2}}H(x,y)dxdy.
\end{equation*}%
And we complete the proof by applying Lemma \ref{lemg1}.

\bigskip

\noindent \textbf{Remark }: We used the proof of Yu(1993) here.

\begin{theorem}
\label{theo1} \label{theog2} Let $X_{1},X_{2},...,X_{n}$ be associated, then
we have for all $t=(t_{1},...,t_{n})\in \mathbb{R}^{k},$%
\begin{equation}
\left\vert \psi _{_{(X_{1},X_{2},...,X_{n})}}(t)-\prod\limits_{i=1}^{n}\psi
_{_{X_{i}}}(t_{i})\right\vert \leq \frac{1}{2} \sum_{1\leq i\neq j\leq
n}\left\vert t_{i}t_{j}\right\vert \left\vert Cov(X_{i},X_{j})\right\vert .
\label{decomp}
\end{equation}
\end{theorem}

\bigskip

\bigskip \noindent \textbf{Proof :} First, we prove this for $n=2.$ Use the
Newman inequality in Lemma \ref{lemg2}. Let $X$ and $Y$ be two associated
random variables. For $(s,t)\in R^{2},$ put $U=f(X)=:e^{isX}$ and $%
V=g(Y)=:e^{itY}.$ We have%
\begin{equation*}
Cov(U,V)=E(e^{(isX+tY)})-E(e^{isX})E(e^{itY})=\psi _{(X,Y)}(s,t)-\psi
_{X}(s)\psi _{Y}(t).
\end{equation*}

\bigskip \noindent But Lemma \ref{lemg2} implies 
\begin{equation*}
\left\vert Cov(U,V)\right\vert =\left\vert Cov(f(X),g(Y))\right\vert \leq
\left\vert st\right\vert \left\Vert f^{\prime }\right\Vert_{\infty}
\left\Vert g^{\prime }\right\Vert \left\vert_{\infty} Cov(X,Y)\right\vert
=\left\vert st\right\vert \left\vert Cov(X,Y)\right\vert .
\end{equation*}

\begin{equation*}
=\frac{1}{2} \left\vert st\right\vert \left\vert (Cov(X,Y)+cov(Y,X))
\right\vert .
\end{equation*}

\bigskip \noindent And ($\ref{decomp}$) is valid for $n=2.$ Now we proceed
by induction and suppose that \ref{decomp} is true up to $n.$ Consider
associated random variables $X_{1},X_{2},...,X_{n+1}$ and let $%
t=(t_{1},...,t_{n+1})\in R^{n+1}.$ If all the $t_{i}$ are nonnegative, we
have $U=t_{1}X_{1}+...+t_{n}X_{n}$ and $V=X_{n+1}$ are associated. We have%
\begin{equation*}
\psi _{_{(X_{1},X_{2},...,X_{n}+1)}}(t)=\psi _{_{(U,V)}}(1,t_{n+1})\text{
and \ }\psi _{_{U}}(1)=\psi _{_{(X_{1},X_{2},...,X_{n})}}(t_{1},...,t_{n}).
\end{equation*}

\bigskip \noindent By the induction hypothesis, we have%
\begin{equation}
\left\vert \psi _{_{(X_{1},X_{2},...,X_{n}+1)}}(t)-\psi
_{_{(X_{1},X_{2},...,X_{n})}}(t_{1},...,t_{n})\psi
_{X_{n+1}}(t_{n+1})\right\vert  \label{d2}
\end{equation}%
\begin{equation*}
\leq \left\vert t_{n+1}\right\vert \left\vert
cov(X_{n+1},t_{1}X_{1}+...+t_{n}X_{n})\right\vert
\end{equation*}%
\begin{equation*}
\leq \frac{1}{2} \sum_{j=1}^{n}\left\vert t_{i}t_{n+1}\right\vert \left\vert
cov(X_{n+1},X_{i})\right\vert .
\end{equation*}

\bigskip \noindent Next 
\begin{equation*}
\left\vert \psi
_{_{(X_{1},X_{2},...,X_{n}+1)}}(t)-\prod\limits_{i=1}^{n+1}\psi
_{_{X_{i}}}(t_{i})\right\vert
\end{equation*}%
\begin{equation*}
\leq \left\vert \psi _{_{(X_{1},X_{2},...,X_{n}+1)}}(t)-\psi
_{_{(X_{1},X_{2},...,X_{n})}}(t_{1},...,t_{n})\psi
_{X_{n+1}}(t_{n+1})\right\vert
\end{equation*}%
\begin{equation*}
+\left\vert \psi _{_{(X_{1},X_{2},...,X_{n})}}(t_{1},...,t_{n})\psi
_{X_{n+1}}(t_{n+1})-\prod\limits_{i=1}^{n+1}\psi
_{_{X_{i}}}(t_{i})\right\vert .
\end{equation*}

\bigskip \noindent The first term in the right side member is bounded as in (%
\ref{d2}). The second term is bounded, due to the induction hypothesis, by%
\begin{equation*}
\left\vert \psi _{X_{n+1}}(t_{n+1})\right\vert \left\vert \psi
_{_{(X_{1},X_{2},...,X_{n})}}(t_{1},...,t_{n})-\prod\limits_{i=1}^{n}\psi
_{_{X_{i}}}(t_{i})\right\vert
\end{equation*}%
\begin{equation*}
=\left\vert \psi
_{_{(X_{1},X_{2},...,X_{n})}}(t_{1},...,t_{n})-\prod\limits_{i=1}^{n}\psi
_{_{X_{i}}}(t_{i})\right\vert
\end{equation*}%
\begin{equation}
\leq \frac{1}{2} \sum_{1\leq i\neq j\leq n}\left\vert t_{i}t_{j}\right\vert
\left\vert cov(X_{i},X_{j})\right\vert .  \label{d3}
\end{equation}

\bigskip \noindent By putting (\ref{d2}) and (\ref{d3}) together, we get
that (\ref{decomp}) is valid. By re-arranging the $t_{i},$ we observe that
we have proved( \ref{decomp}) for $n=3$. if at least $n$ of the $t_{i}$ are
nonnegative. Also, if at least $n$ of them are nonpositive, we consider the
sequence $-X_{1},...,-X_{n+1}$ that is also associated and get the same
conclusion. This means that (\ref{decomp}) is true. It remains the case
where exactely $p$ of the $t_{i}$ are nonnegative with $2\leq p\leq n-2.$ By
re-arranging the $t_{i}$ if necessary, we may consider that $t_{i}\geq 0$
for $1\leq i\leq p$ and $t_{i}<0$ for $i>p.$ Now, by putting $%
U=t_{1}X_{1}+...+t_{p}X_{p}$ and $U=t_{p+1}X_{p+1}+...+t_{n+1}X_{n+1}.$
Since $U$ et $-V$ are associated and since%
\begin{equation*}
\psi _{_{(X_{1},X_{2},...,X_{n}+1)}}(t)=\psi _{(U,-V)}(1,-1),
\end{equation*}

\bigskip \noindent we have by the induction hypothesis%
\begin{equation}
\left\vert \psi _{_{(X_{1},X_{2},...,X_{n}+1)}}(t)-\psi _{U}(1)\psi
_{-V}(-1)\right\vert \leq \frac{1}{2} \left\vert Cov(U,-V)\right\vert \leq 
\frac{1}{2} \sum_{i=1}^{p}\sum_{j=p+1}^{n+1} \left\vert t_i t_j\right\vert
\left\vert cov(X_{i},X_{j})\right\vert  \label{d3a}
\end{equation}

\bigskip \noindent Now use 
\begin{equation}
\left\vert \psi
_{_{(X_{1},X_{2},...,X_{n}+1)}}(t)-\prod\limits_{i=1}^{n+1}\psi
_{_{X_{i}}}(t_{i})\right\vert \leq \left\vert \psi
_{_{(X_{1},X_{2},...,X_{n}+1)}}(t)-\psi _{U}(1)\psi _{-V}(-1)\right\vert
\label{d4}
\end{equation}%
\begin{equation*}
+\left\vert \psi _{U}(1)\psi _{-V}(-1)-\psi
_{U}(1)\prod\limits_{i=p+1}^{n+1}\psi _{_{X_{i}}}(t_{i})\right\vert
\end{equation*}%
\begin{equation*}
\leq \left\vert \psi _{_{(X_{1},X_{2},...,X_{n}+1)}}(t)-\psi _{U}(1)\psi
_{-V}(-1)\right\vert +\left\vert \psi _{U}(1)\psi
_{-V}(-1)-\prod\limits_{i=1}^{p}\psi _{_{X_{i}}}(t_{i})\psi
_{-V}(-1)(t_{i})\right\vert
\end{equation*}%
\begin{equation*}
+\left\vert \prod\limits_{i=1}^{p}\psi _{_{X_{i}}}(t_{i})\psi
_{-V}(-1)-\prod\limits_{i=1}^{n+1}\psi _{_{X_{i}}}(t_{i})\right\vert
\end{equation*}

\bigskip \noindent The first term already handled in (\ref{d4}). The second
term is bounded as follows%
\begin{equation*}
\left\vert \psi _{U}(1)\psi _{-V}(-1)-\prod\limits_{i=1}^{p}\psi
_{_{X_{i}}}(t_{i})\psi _{-V}(-1)(t_{i})\right\vert =\left\vert \psi
_{-V}(-1)(t_{i})\right\vert \times \left\vert \psi
_{U}(1)-\prod\limits_{i=1}^{p}\psi _{_{X_{i}}}(t_{i})\right\vert
\end{equation*}%
\begin{equation*}
\leq \left\vert \psi _{U}(1)-\prod\limits_{i=1}^{p}\psi
_{_{X_{i}}}(t_{i})\right\vert =\left\vert \psi
_{_{(X_{1},X_{2},...,Xp)}}(t_{1},...,t_{p})-\prod\limits_{i=1}^{p}\psi
_{_{X_{i}}}(t_{i})\right\vert
\end{equation*}%
\begin{equation}
\leq \frac{1}{2} \sum_{1\leq i\neq j\leq p}^{p} \left\vert t_i t_j
\right\vert \left\vert cov(X_{i},X_{j})\right\vert.  \label{d5}
\end{equation}

\bigskip \noindent where we used the induction hypothesis in the last
formula. The last term is%
\begin{equation*}
\left\vert \prod\limits_{i=1}^{p}\psi _{_{X_{i}}}(t_{i})\psi
_{-V}(-1)-\prod\limits_{i=1}^{n+1}\psi _{_{X_{i}}}(t_{i})\right\vert
=\left\vert \prod\limits_{i=1}^{p}\psi _{_{X_{i}}}(t_{i})\psi
_{-V}(-1)-\prod\limits_{i=1}^{n+1}\psi _{_{X_{i}}}(t_{i})\right\vert
\end{equation*}%
\begin{equation*}
\leq \left\vert \prod\limits_{i=1}^{p}\psi _{_{X_{i}}}(t_{i})\right\vert
\times \left\vert \psi _{-V}(-1)-\prod\limits_{i=p+1}^{n+1}\psi
_{_{X_{i}}}(t_{i})\right\vert
\end{equation*}%
\begin{equation*}
\leq \left\vert \psi
_{_{(X_{p+1},...,X_{n+1})}}(t_{p+1},...,t_{n+1})-\prod\limits_{i=p+1}^{n+1}%
\psi _{_{X_{i}}}(t_{i})\right\vert
\end{equation*}%
\begin{equation}
\leq \frac{1}{2} \sum_{p+1\leq i\neq j\leq n+1} \left\vert t_i
t_j\right\vert \left\vert cov(X_{i},X_{j})\right\vert,  \label{d6}
\end{equation}

\bigskip \noindent where we used again the induction hypothesis. We complete
the proof by putting (\ref{d3a}), (\ref{d4}), (\ref{d6}) and (\ref{d5})
together, we arrive at the result (\ref{decomp}).

\section{Central limit theorem for a strictly stationary and associated
sequence}

\noindent In this section, we provide all the details of the sharpest result
in this topic by \cite{newmanwright}. This came as a
concluding paper for a series of papers by Newman.\newline

\noindent We present here all the materials used in the proof of Newman and
Wright in a detailed writing that makes it better understandable by a broad
public.\newline

\noindent First, we have this simple lemma.

\begin{lemma}
\label{lemg3} Let $X$ and $Y$ be finite variance random variables such that 
\begin{equation}
E(X,Y1_{(Y\leq 0)})\geq 0.  \label{c1}
\end{equation}%
Then, we have 
\begin{equation}
\mathbb{E}[(\max (X,X+Y))^{2}]\leq E(X+Y)^{2}.  \label{c2}
\end{equation}
\end{lemma}

\bigskip \noindent If $X$ and $Y$ are associated and $X$ is mean zero, then (%
\ref{c1}) holds and (\ref{c2}) is true.\newline

\noindent \textbf{Proof.} We have%
\begin{equation*}
\max (X,X+Y)^{2}=\left\{ X1_{(Y\leq 0)}+(X+Y)1_{(Y>0)}\right\} ^{2}
\end{equation*}%
\begin{equation*}
=X^{2}1_{(Y\leq 0)}+(X+Y)^{2}1_{(Y>0)}=X^{2}1_{(Y\leq
0)}+(X^{2}+Y^{2}+2XY)1_{(Y>0)}
\end{equation*}%
\begin{equation*}
=X^{2}+Y^{2}-Y^{2}1_{(Y\leq 0)}+2(XY)1_{(Y>0)}
\end{equation*}

\qquad 
\begin{equation*}
=X^{2}+Y^{2}+2XY-2XY1_{(Y\leq 0)}-Y^{2}1_{(Y\leq 0)}
\end{equation*}

\begin{equation*}
=\left( X+Y\right) ^{2}-2XY1_{(Y\leq 0)}-Y^{2}1_{(Y\leq 0)}
\end{equation*}

\bigskip \noindent We get the desired result whenever%
\begin{equation*}
E(XY1_{(Y\leq 0)})=Cov(X,Y1_{(Y\leq 0)})\geq 0
\end{equation*}%
Now if $X$ \ and $Y$ are associated, we have%
\begin{equation*}
XY1_{(Y\leq 0)}=(-X)(-Y)1_{(-Y\geq 0)}.
\end{equation*}

\bigskip \noindent Since $(-X)$ and $(-Y)$ are associated too and $%
1_{(-Y\geq 0)}$ is a nondecreasing function of $(-Y)$, and reminding that $X$
is mean zero, we get that 
\begin{equation*}
E(XY1_{(Y\leq 0)})=E((-X)(-Y)1_{(-Y\geq 0)})=Cov((-X),(-Y)1_{(-Y\geq
0)})\geq 0.
\end{equation*}

\begin{theorem}[Maximal inequality of Newman and Wright]
\label{theo2} Let $X_{1},X_{2},\cdots ,X_{n}$ be associated, mean zero,
finite variance, random variables and $M_{n}=\max (S_{1},S_{2},\cdots
,S_{n}) $ where $S_{n}=X_{1}+X_{2}+\cdots +X_{n}$, we have 
\begin{equation}
\mathbb{E}(M_{n}^{2})\leq V(S_{n}).  \label{nw}
\end{equation}
\end{theorem}

\noindent \textbf{Proof.} Let us prove (\ref{nw}) by induction. It is
obviously true \ for $n=1$ and for $n=2$ by Lemma \ref{lemg3}. Let us
suppose that it is true for $j,2\leq j<n.$ By putting $%
L_{j}=X_{2}+...+X_{j}, $ $j\geq 2,$we have

\begin{equation*}
M_{n}=\max (X_{1},X_{1}+L_{2},...,X_{1}+L_{n})=X_{1}+\max(0,L_{2},...,L_{n}).
\end{equation*}%
But%
\begin{equation*}
\max (X_{1},X_{1}+\max (L_{2},...,L_{n}))=X_{1}+\max (0,\max
(L_{2},...,L_{n}))
\end{equation*}%
We obviously have%
\begin{equation*}
\max (0,\max (L_{2},...,L_{n}))=\max (0,L_{2},...,L_{n}).
\end{equation*}

\bigskip \noindent Then 
\begin{equation*}
\mathbb{E}M_{n}^{2}=E\max (X_{1},X_{1}+\max (L_{2},...,L_{n}))^{2}
\end{equation*}

\bigskip \noindent Since $X_{1}$ and $\max (L_{2},...,L_{n})$ are associated
and $X_{1}$ is mean zero, then use Lemma \ref{lemg3} to get%
\begin{equation*}
EM_{n}^{2}=E\max (X_{1},X_{1}+\max (L_{2},...,L_{n}))^{2}\leq
EX_{1}^{2}+E\max (L_{2},...,L_{n})^{2}
\end{equation*}

\bigskip \noindent And then, apply (\ref{nw}) on $\mathbb{E}\max
(L_{2},...,L_{n})^{2}$ for $(n-1)$ mean zero associated rv's to have%
\begin{equation*}
\mathbb{E}\max (L_{2},...,L_{n})^{2}\leq EX_{2}^{2}+...+X_{n}^{2}.
\end{equation*}%
We conclude that%
\begin{equation*}
EM_{n}^{2}\leq \mathbb{E}X_{1}^{2}+\mathbb{E}X_{2}^{2}+...+\mathbb{E}%
X_{n}^{2}.
\end{equation*}

\bigskip

\begin{lemma}
Let $X_{1},X_{2},\cdots ,X_{n}$ be a second-order stationary sequence with $%
\sigma ^{2}=V(X_{1})+2\sum_{j=2}^{\infty }|Cov(X_{1},X_{j})|<\infty ,$ then 
\begin{equation*}
V\left( \dfrac{S_{n}}{\sqrt{n}}\right) \rightarrow \sigma
^{2}=V(X_{1})+2\sum_{j=2}^{\infty }Cov(X_{1},X_{j}).
\end{equation*}
\end{lemma}

\noindent\textbf{Proof.} We have 
\begin{equation*}
\alpha _{n}=V\left( \dfrac{S_{n}}{\sqrt{n}}\right) =\frac{1}{n}\left\{
\sum_{j=1}^{n}V(X_{i})+\sum_{1\leq i\neq j\leq n}Cov(X_{i},X_{j})\right\} .
\end{equation*}

\bigskip \noindent By stationarity, we have%
\begin{equation*}
V\left( \dfrac{S_{n}}{\sqrt{n}}\right) =V(X_{1})+\frac{2}{n}\sum_{1\leq
i<j\leq n}Cov(X_{i},X_{j})
\end{equation*}%
\begin{equation*}
=V(X_{1})+\frac{2}{n}\sum_{j=2}^{n}(n-j+1)Cov(X_{1},X_{j}).
\end{equation*}%
Let $\epsilon >0$. Since $\sum_{j=2}^{\infty }Cov(X_{1},X_{j}) < +\infty$,
there exists $K>0$ such that for any $k\geq K,$ 
\begin{equation*}
\sum_{j\geq k+1}Cov(X_{1},X_{j})<\epsilon .
\end{equation*}

\bigskip \noindent We fix that $k\geq K$ and write, 
\begin{equation*}
\alpha _{n}=V(X_{1})+2\left[ \sum_{j=2}^{k}\left( 1-\dfrac{j-1}{n}\right)
Cov(X_{1},X_{j})+\sum_{j=k+1}^{n}\left( 1-\dfrac{j-1}{n}\right)
Cov(X_{1},X_{j})\right]
\end{equation*}

\bigskip \noindent and observe that 
\begin{equation*}
\left\vert \alpha _{n}-V(X_{1})-2\sum_{j=2}^{k}\left( 1-\dfrac{j-1}{n}%
\right) Cov(X_{1},X_{j})\right\vert \leq 2\epsilon .
\end{equation*}

\bigskip \noindent Thus, we get 
\begin{equation*}
\lim \inf V(X_{1})+2\sum_{j=2}^{k}\left( 1-\dfrac{j-1}{n}\right)
Cov(X_{1},X_{j})-2\epsilon \leq \lim \inf \alpha _{n}
\end{equation*}%
\begin{equation*}
\leq \lim \sup \alpha _{n}\leq \lim \sup V(X_{1})+2\sum_{j=2}^{k}\left( 1-%
\dfrac{j-1}{n}\right) Cov(X_{1},X_{j})+2\epsilon .
\end{equation*}

\bigskip \noindent Therefore, for any $k\geq K,$ 
\begin{equation*}
V(X_{1})+2\sum_{j=2}^{k}Cov(X_{1},X_{j})-2\epsilon \leq \lim \inf \alpha
_{n}\leq \lim \sup \alpha _{n}
\end{equation*}%
\begin{equation*}
\leq V(X_{1})+2\sum_{j=2}^{k}Cov(X_{1},X_{j})+2\epsilon .
\end{equation*}%
We finish the proof by letting $k\rightarrow \infty $ and next by \ letting $%
\epsilon \rightarrow 0.$

\begin{theorem}
Let $X_{1},X_{2},\cdots ,X_{m}$ be a strictly stationary, mean zero,
associated random variables such that 
\begin{equation*}
\sigma ^{2}=V(X_{1})+2\sum_{j=2}^{+\infty}Cov(X_{1},X_{j})<\infty ,
\end{equation*}%
then 
\begin{equation*}
\dfrac{S_{n}}{\sqrt{n}}=\dfrac{X_{1}+X_{2}+\cdots +X_{n}}{\sqrt{n}}%
\rightarrow N(0,\sigma ^{2})\ as\ n\rightarrow \infty
\end{equation*}
\end{theorem}

\noindent \textbf{Proof.} Let us fix $\ell >1$ an integer and let us set $m=[%
\frac{n}{\ell }]$, that is $m\ell \leq n\leq m\ell +\ell .$ Let us define $%
\Psi _{n}(r)=\mathbb{E}(e^{irS_{n}/\sqrt{n}}),$ $r\in \mathbb{R}$. \noindent
First, we have for $r\in \mathbb{R}$, 
\begin{equation*}
|\Psi _{n}(r)-\Psi _{m\ell }(r)|=|\mathbb{E}(e^{irS_{n}/\sqrt{n}})-\mathbb{E}%
(e^{irS_{m\ell }/\sqrt{m\ell }})|
\end{equation*}%
\begin{equation*}
=\left\vert \mathbb{E}\left[ e^{irS_{m\ell }/\sqrt{m\ell }}\left( e^{ir\left[
(S_{n}/\sqrt{n})-(S_{m\ell }/\sqrt{m\ell })\right] }-1\right) \right]
\right\vert 
\end{equation*}%
\begin{equation}
\leq \mathbb{E}\left\vert e^{ir\left( \frac{S_{n}}{\sqrt{n}}-\frac{S_{m\ell }%
}{\sqrt{m\ell }}\right) }-1\right\vert .  \label{b}
\end{equation}%
But for any $x\in \mathbb{R}$, 
\begin{equation*}
|e^{ix}-1|=|(\cos x-1)+i\sin x|=|2\sin \frac{x}{2}|\leq |x|.
\end{equation*}

\bigskip Thus the second member of $(\ref{b})$ is, by the Cauchy-Schwarz's
inequality, bounded by 
\begin{equation*}
|r|\mathbb{E}\left\vert \frac{S_{n}}{\sqrt{n}}-\frac{S_{m\ell }}{\sqrt{m\ell 
}}\right\vert \leq |r|V\left( \frac{S_{n}}{\sqrt{n}}-\frac{S_{m\ell }}{\sqrt{%
m\ell }}\right) ^{\frac{1}{2}}.
\end{equation*}

\bigskip \noindent Let us compute the quantity between brackets for fixed $%
\ell $ and $n\rightarrow \infty $ $(m\rightarrow \infty )$, we get 
\begin{equation*}
\frac{S_{n}}{\sqrt{n}}-\frac{S_{m\ell }}{\sqrt{m\ell }}=\frac{S_{n}}{\sqrt{n}%
}-\frac{S_{m\ell }}{\sqrt{n}}+\frac{S_{m\ell }}{\sqrt{n}}-\frac{S_{m\ell }}{%
\sqrt{m\ell }}
\end{equation*}%
\begin{equation*}
=\frac{S_{n}-S_{m\ell }}{\sqrt{n}}-\frac{\sqrt{n}-\sqrt{m\ell }}{\sqrt{%
nm\ell }}S_{m\ell }
\end{equation*}%
and 
\begin{equation*}
\delta _{m,\ell }=V\left( \frac{S_{n}}{\sqrt{n}}-\frac{S_{m\ell }}{\sqrt{%
m\ell }}\right) =V\left( \frac{S_{n}-S_{m\ell }}{\sqrt{n}}\right) +\left( 
\frac{\sqrt{n}-\sqrt{m\ell }}{\sqrt{n}}\right) ^{2}V\left( \frac{S_{m\ell }}{%
\sqrt{m\ell }}\right) 
\end{equation*}%
\begin{equation*}
-2\frac{\sqrt{n}-\sqrt{m\ell }}{\sqrt{nm\ell }}Cov(S_{n}-S_{m\ell },S_{m\ell
}).
\end{equation*}%
$Cov(S_{n}-S_{m\ell },S_{m\ell })\geq 0$ by association. Thus 
\begin{equation*}
\delta _{m,\ell }\leq V\left( \frac{S_{n-m\ell }}{\sqrt{n}}\right) +\left( 
\frac{\sqrt{n}-\sqrt{m\ell }}{\sqrt{n}}\right) ^{2}V\left( \frac{S_{m\ell }}{%
\sqrt{m\ell }}\right) .
\end{equation*}%
Since $0\leq n-m\ell \leq \ell $, and $Cov(X_{1},X_{j})\geq 0$ by
association, 
\begin{equation*}
V(S_{n-m\ell })=\sum_{i=1}^{n-m\ell }V(X_{i})+\sum_{1\leq i\neq j\leq
n-m\ell }Cov(X_{i},X_{j})
\end{equation*}%
\begin{equation*}
\leq \sum_{i=1}^{\ell }V(X_{i})+\sum_{1\leq i\neq j\leq \ell
}Cov(X_{i},X_{j})=A(\ell ).
\end{equation*}

\bigskip \noindent Further, $m\ell \leq n\leq (m+1)\ell $ implies 
\begin{equation*}
0\leq \frac{\sqrt{n}-\sqrt{m\ell }}{\sqrt{n}}\leq \left( 1-\sqrt{\frac{m\ell 
}{n}}\right) \rightarrow 0\text{ as }n\rightarrow +\infty .
\end{equation*}%
Then when $m\rightarrow \infty $ $(n\rightarrow \infty )$ 
\begin{equation*}
V\left( \frac{S_{m\ell }}{\sqrt{m\ell }}\right) \rightarrow
V(X_{1})+2\sum_{j=2}^{\infty }Cov(X_{1},X_{j})<\infty 
\end{equation*}%
and 
\begin{equation*}
\delta _{m,\ell }\leq \frac{A(\ell )}{n}+\left( 1-\sqrt{\frac{m\ell }{n}}%
\right) ^{2}V\left( \frac{S_{m\ell }}{\sqrt{m\ell }}\right) \rightarrow 0
\end{equation*}%
for fixed $\ell $, $n\rightarrow \infty $, we get 
\begin{equation*}
|\Psi _{n}(r)-\Psi _{m\ell }(r)|\rightarrow 0.
\end{equation*}

\bigskip \noindent Now, let us set $Y_{j}=(S_{j\ell }-S_{\ell (j-1)})/\sqrt{%
\ell }$, for a fixed $\ell $. By strict stationarity, the $Y_{j}^{\prime }s$
are associated and identically distributed. Let $\Psi_{\ell} $ be the common
characteristic function of $Y_{1},\cdots ,Y_{m}$. Furthermore 
\begin{equation*}
\frac{S_{m\ell }}{\sqrt{m\ell }}=\frac{1}{\sqrt{m}\sqrt{\ell }}%
\sum_{j=1}^{m}(S_{j\ell }-S_{\ell (j-1)})=\frac{1}{\sqrt{m}}%
\sum_{j=1}^{m}Y_{j}.
\end{equation*}

\bigskip \noindent According to the Newman's Theorem (see Theorem \ref{theo1}%
) 
\begin{equation*}
\left\vert \Psi _{m\ell }(r)-\left( \Psi _{\ell }\left( \frac{r}{\sqrt{m}}%
\right) \right) ^{m}\right\vert \leq \frac{r^{2}}{2m}\sum_{1\leq j\neq k\leq
m}Cov(Y_{j},Y_{k}),
\end{equation*}%
and we know that 
\begin{equation*}
V\left( \sum_{j=1}^{m}Y_{j}\right) =\sum_{j=1}^{m}V(Y_{j})+\sum_{1\leq j\neq
k\leq m}Cov(Y_{j},Y_{k}).
\end{equation*}

\bigskip \noindent Thus, by using the stationarity again, we get 
\begin{equation*}
\frac{1}{m}\sum_{1\leq j\neq k\leq m}Cov(Y_{j},Y_{k})=\frac{1}{m}V\left(
\sum_{j=1}^{m}Y_{j}\right) -\frac{1}{m}\sum_{j=1}^{m}V(Y_{j})
\end{equation*}%
\begin{equation*}
=V\left( \frac{1}{\sqrt{m}}\sum_{j=1}^{m}Y_{j}\right) -\frac{1}{m}%
\sum_{j=1}^{m}V\left( Y_{j}\right) 
\end{equation*}%
\begin{equation*}
V\left( \frac{S_{m\ell }}{\sqrt{m\ell }}\right) -V\left( \frac{S_{\ell }}{%
\sqrt{\ell }}\right) =\sigma _{m\ell }^{2}-\sigma _{\ell }^{2},
\end{equation*}%
where for any $p\geq 2,$ 
\begin{equation*}
\sigma _{p}^{2}=\frac{1}{p}\sum_{i=1}^{p}V(Y_{i})+\frac{1}{p}\sum_{1\leq
i\neq j\leq p}Cov(Y_{i},Y_{j})
\end{equation*}

\noindent Now, when $m\rightarrow \infty $, $\sigma _{m\ell }^{2}\rightarrow \sigma
^{2}$ and 
\begin{equation*}
\left( \Psi _{\ell }\left( \frac{r}{\sqrt{m}}\right) \right) ^{m}\rightarrow
e^{-\sigma _{\ell }^{2}r^{2}/2},
\end{equation*}

\bigskip \noindent where $\sigma _{\ell }^{2}$ is the common variance of $%
Y_{j}^{\prime }s$, 
\begin{equation*}
\sigma _{\ell }^{2}=\sum_{i=1}^{\ell }V(X_{i})+\frac{1}{\ell }\sum_{1\leq
i\neq j\leq m}Cov(X_{i},X_{j}).
\end{equation*}

\noindent Then it comes out that 
\begin{equation*}
\varlimsup \left\vert \Psi _{m\ell }(r)-e^{-\sigma _{\ell
}^{2}r^{2}/2}\right\vert \leq \frac{r^{2}}{2}(\sigma ^{2}-\sigma _{\ell
}^{2}).
\end{equation*}

\noindent We complete the proof by letting $\ell \rightarrow \infty .$ Thus $\sigma
_{\ell }^{2}-\sigma ^{2}\rightarrow 0$ and we get 
\begin{equation*}
\lim_{n\rightarrow \infty }\left\vert \Psi _{n}(r)-e^{-\sigma
^{2}r^{2}/2}\right\vert =0.
\end{equation*}

\noindent \textbf{Remark}. We finish this exposition by these important facts. A number of CLT's and invariance principles are available in the literature for strictly stationary sequences of associated random variables and not stationary ones. The most general CLT seems to be the one provided by \cite{cox} for arbitrary associated rv's satisfying a number of moment conditions. \cite{burton} and \cite{dabro}) considered weakly associated random variables to establish invariance principle in the lines of \cite{newmanwright}, as well as Berry-Essen-type results and functional laws of Iterated Logarithm (LIL). But almost all these results use the original adaptation of the original method of Newman we have described here.

\end{document}